# A Novel Antifragility Measure Based on Satisfaction and Its Application to Random and Biological Boolean Networks

Omar K. Pineda[1,2,†], Hyobin Kim[2,†], Carlos Gershenson[2,3,4,*]

[1] Posgrado en Ciencia e Ingeniería de la Computación, Universidad Nacional Autónoma de México, 04510 CDMX, México.

[2] Centro de Ciencias de la Complejidad, Universidad Nacional Autónoma de México, 04510 CDMX, México.

[3] Instituto de Investigaciones en Matemáticas Aplicadas y en Sistemas, Universidad Nacional Autónoma de México, 04510 CDMX, México.

[4] ITMO University, St. Petersburg, 199034, Russian Federation.

[†] These authors contributed equally to this work.

*Correspondence should be addressed to Carlos Gershenson cgg@unam.mx.

## Abstract

Antifragility is a property that enhances the capability of a system in response to external perturbations. Although the concept has been applied in many areas, a practical measure of antifragility has not been developed yet. Here we propose a simply calculable measure of antifragility, based on the change of "satisfaction" before and after adding perturbations, and apply it to random Boolean networks (RBNs). Using the measure, we found that ordered RBNs are the most antifragile. Also, we demonstrated that seven biological systems are antifragile. Our measure and results can be used in various applications of Boolean networks (BNs) including creating antifragile engineering systems, identifying the genetic mechanism of antifragile biological systems, and developing new treatment strategies for various diseases.

## Introduction

Antifragility suggested by Taleb is defined as a property to enhance the capability of a system in response to external stressors [1]. It is beyond resilience or robustness. While the resilient/robust systems resist stress and stay the same, antifragile systems not only withstand stress but also benefit from it. The immune system is a representative example of antifragile systems. When exposed to diverse germs at an early age, our immune system strengthens and thus overcomes new diseases in the future.

The concept of antifragility has been actively applied in numerous areas such as risk analysis [2, 3], physics [4], molecular biology [5, 6], transportation planning [7, 8], engineering [9, 10, 11], aerospace and computer science [12-15]. However, a practical measure of antifragility has not been developed yet. Here we propose a novel measure for antifragility based on the change of complexity. We use random Boolean networks (RBNs) as a case study to illustrate our measure. We quantify the complexity by assessing the extent of how much the node states of a RBN are maintained and changed during state transitions. We perturb the network, flipping the node states with the structure of the network fixed. Calculating the variation of the complexity in the network before and after adding the perturbations, we measure antifragility.

BNs have a wide range of applications from biochemical systems [16-20], to economic systems [21]; from social networks, [22, 23] to robots [24]. Our antifragility measure can be utilized in various applications of BNs. For instance, one could create antifragile engineered systems or identify the genetic mechanisms of antifragile biological systems.

The rest of our article is structured as follows. In the section of "Measurement of Antifragility in RBNs", we describe RBNs, complexity of RBNs, perturbations to RBNs, and how to assess antifragility in RBNs. In the section "Experiments", methods and parameter setting for simulations are explained. In the section of "Results and Discussion", the results of the antifragility of RBNs and several biological BNs are presented and analyzed. The section of "Conclusions" summarizes and closes the article.

## Measurement of Antifragility in RBNs

**Random Boolean Networks**

RBNs were proposed as models of gene regulatory networks by Kauffman [25, 26]. A RBN consists of $N$ nodes representing genes. Each node can take either 0 (off, inhibited) or 1 (on, activated) as its state. The node state is determined by the states of input nodes and Boolean functions assigned to each node. Every node has $K$ input nodes (or input links). Self-inputs are allowed. The links are wired randomly, and the Boolean functions are also randomly assigned. Once the links and the Boolean functions set up, they remain fixed.

In Figure 1(a) and (b), the left plots show how randomly chosen initial states are updated over time. The plots are simulated until $T = 40$. A state space refers to the set of all the possible configurations ($2^N$) and all the transitions among them. Being deterministic, classic RBNs have one and only one successor for each state. In the state space, repeated states are attractors, which can be fixed points or limit cycles. The other states that lead to the attractors are basin of attraction of the attractors.

Depending on the structure of the state space, there are three dynamical regimes in RBNs: ordered, chaotic, and critical. The first two are phases, while the critical regime lies at the phase transition. Ordered dynamics are characterized by the change of few node states, which is related to high stability. Chaotic dynamics are characterized by the change of most node states, which is associated with high variability. Critical dynamics balance the stability of the ordered regime with the variability of the chaotic regime [27, 28]. The dynamical regimes can be varied by $K$. For RBNs with internal homogeneity (*i.e.*, the probability that a gene is activated [29]) $p = 0.5$, $K = 1$ is ordered, $K = 2$ is critical, and $K > 2$ is chaotic, on average [30]. Other properties of RBNs can be used to regulate dynamical regimes [31].

**Complexity of RBNs**

It is well known that complex adaptive systems are equipped with stability and flexibility simultaneously. Here complexity signifies a balance between regularity and change, which allows systems to adapt robustly [26, 32, 33]. From an information viewpoint, the regularity ensures that useful information survives, while the change enables the systems to explore new possibilities essential for adaptability [34]. Living organisms or computer systems need not only stability to survive or to maintain information but also flexibility to evolve and adapt to their environment. Following this concept of complexity, we developed a quantitative measure [27]. Using our previous approach, we can measure the complexity of RBNs. In this study, complexity is presented as quantities computed according to our measure.

The complexity is calculated based on Shannon's information entropy. Its equation is as follows:

$$E_i = -(p_0 \log_2 p_0 + p_1 \log_2 p_1) \quad (1)$$

$$C = 4 \times \bar{E} \times (1 - \bar{E}) \quad (2)$$

where $E_i$ is the "emergence" of node $i$, $p_j$ is the probability that the state of the node is $j$ ($j = 0, 1$) among the states of node $i$ updated at each time step until simulation time $T$, $C$ ($0 \leq C \leq 1$) is the complexity of the network, and $\bar{E}$ ($0 \leq \bar{E} \leq 1$) is the average of the emergence values for all the nodes. Specifically, $p_0$ ($p_1$) is calculated by counting the number of 0s (1s) in node $i$ until simulation time $T$. For example, in the left plot of Figure 1(a), $p_0$ and $p_1$ of the last node are $\frac{2}{40}$ and $\frac{38}{40}$, respectively. Because RBNs are deterministic systems, once initial states are determined, state transitions from them to attractors are also determined. Thus, $\bar{E}$ and $C$ are dependent on initial states.

When $\bar{E}$ is calculated, $p_0$ and $p_1$ in equation (1) should not be confused with $p$ which was mentioned as internal homogeneity in previous section. $p_0$ and $p_1$ are values measured from state transitions. Meanwhile, $p$ is a parameter used to create Boolean functions assigned to each node in a RBN. In the Boolean functions, each value is determined with probability $p$ of being one or probability $1 - p$ of being zero.

$\bar{E}$, $1 - \bar{E}$, and $C$ are time-dependent because they focus on the dynamics of node states. $\bar{E}$ indicates how much new states are produced over time (*i.e.*, change). As the complement of $\bar{E}$, $1 - \bar{E}$ represents how much existing states are maintained (*i.e.*, regularity). $C$ means how successfully the both of them are met. Numerically, $C$ reaches maximum when the emergence $\bar{E}$ is 0.5 ($\bar{E} = 0.5 \rightarrow C = 1$). It is when the expression of any one of the two states is highly probable, *i.e.*, $p_0$ or $p_1 \cong 0.89$ for each node [28, 35]. Meanwhile, $C$ becomes 0 when the two states are evenly distributed ($p_0 = p_1 = 0.5$; $\bar{E} = 1$) or only one state has maximum probability ($p_0$ or $p_1 = 1$; $\bar{E} = 0$).

Figure 2 illustrates a mathematical relation between change $\bar{E}$, regularity $1 - \bar{E}$, and complexity $C$ in RBNs [35]. As seen in the figure, high complexity is achieved when $\bar{E} = 1 - \bar{E}$, which means an optimal balance between keeping and changing the states of the network. For perturbed RBNs, Figure 1(a) shows that the antifragile network maintains original states overall, and simultaneously explores new states by means of perturbations. Figure 1(b) represents that most of the states in the fragile network change with perturbations, which indicates that the network does not maintain information in a noisy environment.

**Network Perturbations**

We express network perturbations due to external stressors as the change of node states in a RBN. We flip the states of $X$ nodes randomly chosen, where the perturbations are added with frequency $O$ during simulation run time $T$. In other words, the perturbations are added whenever the time step $t$ is divisible by $O$ ($t \mod O = 0$). For example, $X = 2$, $O = 3$, and $T = 99$ mean that the states of two nodes randomly chosen in each configuration are flipped every three time steps until the simulation run time becomes 99. By comparing the state transitions of the original network and its perturbed network, we can observe how the perturbations propagate over time (Figure 1).

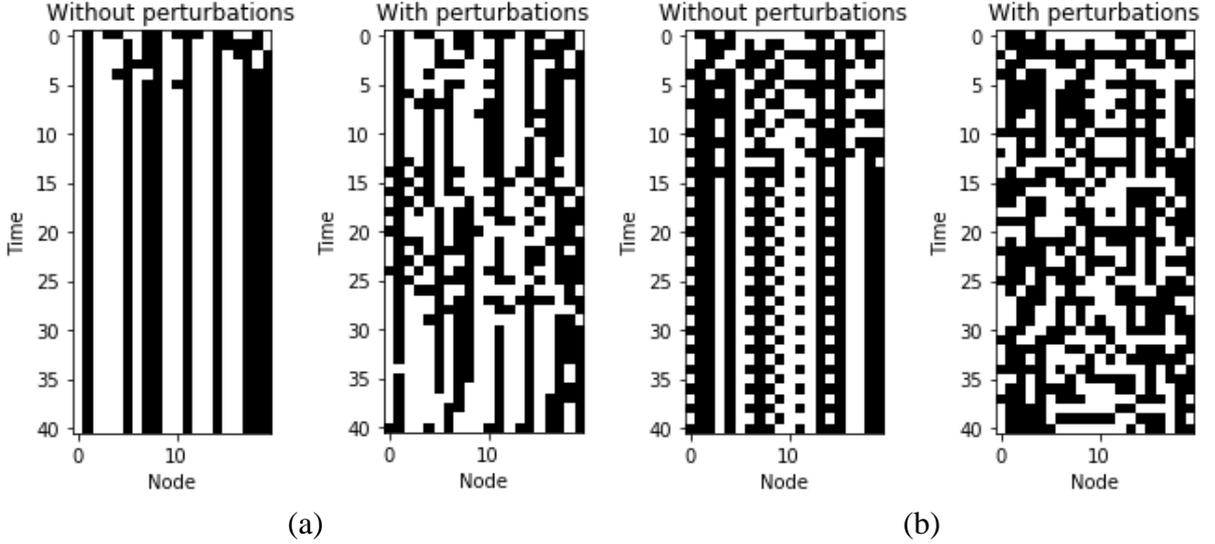

Figure 1: Schematic diagrams showing state transitions of (a) critical and (b) chaotic RBNs with $N = 20$, $X = 2$, and $O = 1$. The left side is the network without perturbations and the right one is the network with perturbations with the same initial states. Each square represents the state of a node (white = 0, black = 1). The state transitions were calculated from the initial states at the top to states at the bottom during $T = 40$. (a) $K = 2$ (critical), $\phi = -0.0958$ (complexity is increased by perturbations: antifragile). (b) $K = 3$ (chaotic), $\phi = 0.0388$ (complexity is decreased by perturbations: fragile).

In our study, the degree of perturbations is defined as follows:

$$\Delta x = \frac{X \times (\frac{T}{O})}{N \times T} \tag{3}$$

where $0 \leq \Delta x \leq 1$.

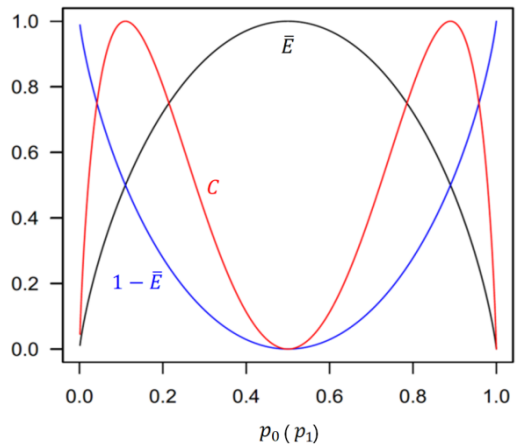

Figure 2: Relationship between change $\bar{E}$, regularity $1 - \bar{E}$, and complexity $C$ in RBNs [35].

## Antifragility of RBNs

We define (*anti*)*fragility* $\oint$ as:

$$\oint = -\Delta\sigma \times \Delta x \qquad (4)$$

where $\Delta\sigma$ is the difference of "satisfaction" before and after perturbations, while $\Delta x$ is the degree of perturbations. To prevent the influence of node size of a network, we calibrate the values of $\oint$ by multiplying $\Delta x$. The satisfaction $\sigma$ is the degree to which the goals of an agent have been achieved [36]. In the context of RBNs, each node of the network can be seen as an agent. We can arbitrarily define their goal as reaching a balance between change and regularity, which is achieved when the nodes have high complexity values. Thus, in RBNs, the satisfaction is measured with complexity. Depending on how the satisfaction changes before and after perturbations, the RBN is classified: fragile, robust, or antifragile.

The satisfaction can be measured differently depending on the particular systems, *e.g.* performance, value, fitness. If the satisfaction is decreased with perturbations, then the system is fragile. If the satisfaction does not change before and after adding perturbations, then the system is robust. If the satisfaction increases with perturbations, then the system is antifragile. Notice that $\Delta\sigma$ and $\Delta x$ should be normalized to the interval [-1,1], [0,1] respectively.

The perturbations $\Delta x$ for RBNs were defined in the previous section. We can define the "satisfaction" of a RBN based on its complexity. Since high complexity offers a balance between robustness and adaptability, we can arbitrarily prefer RBNs with high complexity. Using the complexity measure presented previously, $\Delta\sigma$ is calculated by the following equation:

$$\Delta\sigma = C - C_0 \qquad (5)$$

where $C_0$ is complexity of a network before adding perturbations, and $C$ is complexity of the network after adding perturbations. The same initial states are used at $t = 0$. Because the value of complexity is between 0 and 1, $-1 \leq \Delta\sigma \leq 1$.

Negative values of $\oint$ mean that the RBN is antifragile and positive values mean that the RBN is fragile. Values close to zero indicate that the RBN is robust. As shown in equation (4), $\oint$ has the opposite sign of $\Delta\sigma$. Hence, the negative values of $\oint$ indicate that $C$ is larger than $C_0$ (*i.e.*, the complexity of a system is improved by external perturbations), while the positive values represent that $C_0$ is greater than $C$ (*i.e.*, the complexity is lowered by the perturbations). The value of 0 refers to the complexity does not change before and after perturbations, which indicates that the RBN is robust. Figure 1(a) and 1(b) show the values of $\oint$ calculated from the examples of critical and chaotic RBNs.

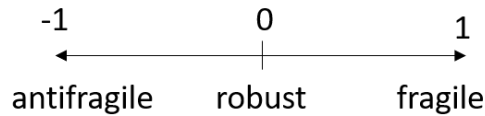

In a RBN, the value of $\oint$ can be different depending on initial states because $\Delta\sigma$ is determined by the states of nodes. Thus, using multiple initial states, we calculate average $\oint$ for a RBN and represent it as a system property.

## Experiments

We performed two sets of experiments: one for RBNs, and the other for biological BNs.

First, to measure antifragility of RBNs, we generated ordered, critical and chaotic RBNs composed of 100 nodes ($K = 1$ (ordered), 2 (critical), 3, 4, 5 (chaotic)) with internal homogeneity $p = 0.5$ [30]. For each RBN, we randomly chose 10 different initial states and then examined their state transitions until simulation time $T = 200$, respectively. For the same RBN taking the same initial states, varying perturbed node size $X$ and perturbation frequency $O$, we obtained the state transitions of the perturbed RBN until $T = 200$. By comparing complexity before and after perturbations, we calculated mean of antifragility for the 10 initial states. The measured values shown in the plots are average calculated from 50 different RBNs per $K$.

Secondly, to measure antifragility of biological BNs, we used the following seven biological network models:

- *CD4+ T cell differentiation and plasticity* [37] ($N = 18$). It is a model representing how *CD4+ T cells* orchestrate immune responses depending on environmental signals and immunological challenges.
- *Mammalian cell-cycle* [38] ($N = 20$). It is a model explaining the mechanism of action of the cell cycle checkpoints in mammalian cells.
- *Cardiac development* [39] ($N = 15$). It is a model referring to how the first heart field (FHF) and second heart field (SHF) are formed by differential expression of transcription and signaling factors during cardiac developmental processes.
- *Metabolic interactions in the gut microbiome* [40] ($N = 12$). It is a model describing interactive host-microbiota metabolic processes.
- *Death receptor signaling* [41] ($N = 28$). It is a model related to the activation of death receptors (TNFR and Fas) that determine either survival or cell death.
- *Arabidopsis thaliana cell-cycle* [42] ($N = 14$). It is a model explaining the mechanism of plant cell-cycle and cell differentiation in *A. thaliana*.
- *Tumor cell invasion and migration* [43] ($N = 32$). It is a model representing the mechanism and interplays between pathways that are involved in the process of metastasis.

For each network, we randomly chose 1000 different initial states and then investigated their state transitions until $T = 200$. Changing $X$ and $O$, we computed antifragility. Specifications of parameters for the simulation follows Table 1. Our simulator for antifragility was implemented in Python[1].

Table 1: Parameter settings for experiments

| Figure | $N$ | $T$ | $X$ | $O$ | # of different networks | # of initial states |
|---|---|---|---|---|---|---|
| 3(a) | 100 | 200 | 1..100 | 1 | 50 | 10 |
| 3(b) | 100 | 200 | 40 | 1..50 | 50 | 10 |
| 4(a) | 100 | 2000 | 0 | 0 | 1000 | 1 |

---

[1] The source code is available at https://github.com/Okarim1/RBN.git

| | | | | | | |
|---|---|---|---|---|---|---|
| 4(b) | 100 | 200 | 1..100 | 1 | 50 | 10 |
| 4(c) | 100 | 200 | 1..100 | 1 | 50 | 10 |
| 5 | N | 200 | 1..N | 1 | 1 | 1000 |
| 6 | 100 | 200 | 1..100 | 1..30 | 50 | 10 |
| 7 | N | 200 | 1..N | 1..20 | 1 | 5000 |

## Results and Discussion

### Antifragility in RBNs

Figure 3 shows average $\oint$ of ordered ($K = 1$), critical ($K = 2$), and chaotic RBNs ($K = 3, 4, 5$) depending on perturbed node size $X$ and perturbation frequency $O$. The ordered and critical RBNs had negative values (antifragility) in certain ranges of $X$ and $O$, while the chaotic RBNs all had zero or positive values in the given ranges. This means that the ordered and critical RBNs can be antifragile if they have the "right" amount of perturbations. However, chaotic RBNs are just robust or fragile against perturbations.

As shown in Figure 3(a), the values of the ordered and critical RBNs were lower than zero and got smaller and smaller as $X$ increased, which indicates that their dynamics change more and more antifragile. However, the values increased beyound certain $X$ values, and even the critical RBNs changed from antifragile to fragile ($X > 20$). From this, we found that neither too large nor too small, but a moderate level of perturbations can induce greater antifragility. These dynamics are similar to the slower-is-faster effect, where a moderate level of speed can lead to better traffic flow rather than that the highest speed of individuals [44].

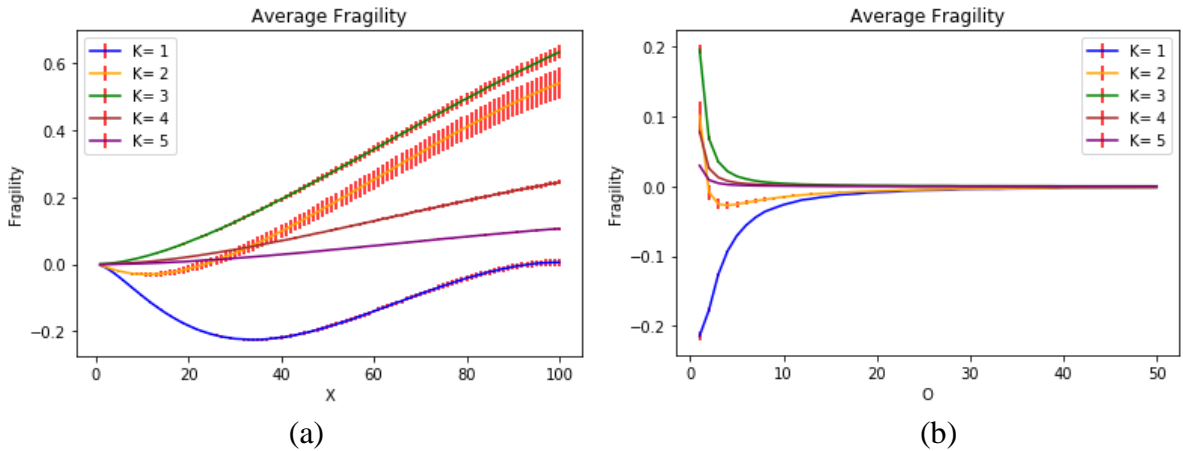

Figure 3: Average $\oint$ of ordered, critical, chaotic RBNs depending on $X$ and $O$. The error bars represent the standard error of measurements for 50 different networks at 10 different initial states ran by 200 steps. (a) $N = 100$ and $O = 1$. (b) $N = 100$ and $X = 40$.

Meanwhile, in Figure 3(b), antifragility of the ordered and critical RBNs decreased overall as $O$ grew (*i.e.*, the period of adding perturbations became longer and longer). Furthermore, all the RBNs were robust in the case of that the perturbations were not added frequently although the perturbed nodes were 40 ($X = 40$). From these results, we found that the more frequently perturbations are added, the more antifragile a system is, particularly for the ordered RBNs. Moreover, how often perturbations are added has a greater effect on antifragility than how many nodes are perturbed. Thus, it is essential that moderate perturbations are added frequently in order to obtain maximal antifragility.

Based on Figure 3, we are able to see that the ordered RBNs are the most antifragile. Figure 4 clearly accounts for the reason. In Figure 4(a), the complexity before adding perturbations was lowest at $K = 1$. However, as shown in Figure 4(b), the complexity after adding perturbations increased most greatly and the value was also largest except for the early range of $X$ at $K = 1$. Therefore, the difference was largest at $K = 1$ (Figure 4(c)), which led the ordered RBNs to be most antifragile.

Our result for complexity before perturbations is the same as previous studies showing that critical RBNs have the most appropriate balance between regularity and change [27, 35, 45]. In Figure 4(a), for low $K$, the complexity was low, which represents that the ordered RBNs have high robustness and few changes. That is, there is few or no information emerging. For high $K$, the complexity was also low, which reflects that the chaotic RBNs have high variability and many changes. Almost all the nodes carry novel emergent information. For medium connectivities ($2 < K < 3$), there was a balance between regularity and change, leading to high complexity. This is consistent with the dynamics of critical RBNs, where criticality is found theoretically at $K = 2$ (when $N \rightarrow \infty$) and for finite systems at $2 < K < 3$ due to a finite-size effect [35].

However, the result is changed by adding perturbations. In Figure 4(b), the ordered RBNs had the biggest complexity excluding the early range of $X$, which means that the ordered RBNs show the optimal balance between regularity and change in the presence of noise. This illustrates that systems can exhibit different properties in accordance with the presence of external stressors. Such phenomenon was recently observed in a neural network as well [46], where neural systems showed different dynamical behaviors depending on the presence/absence of external inputs.

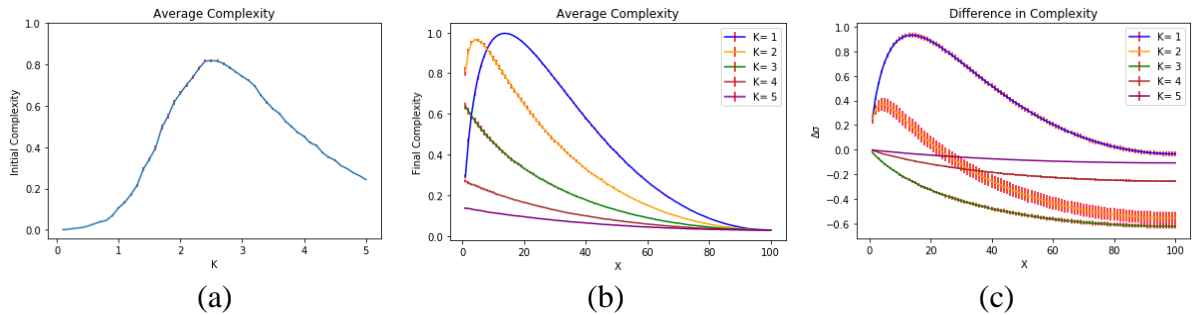

Figure 4: Initial and final complexity for $K = 1, 2, 3, 4, 5$ with $N = 100$. The error bars represent the standard error of measurements for 50 different networks at 10 different initial states run by 200 steps. (a) Complexity before adding perturbations. (b) Complexity after adding perturbations. (c) Difference of complexity before and after perturbations.

**Antifragility in Biological BNs**

Boolean networks have been extensively used as models of genetic or cellular regulation in the fields of computational and systems biology [37-43], because they can capture interesting features of biological systems despite their simplicity. Using seven biological Boolean network models, we measured the values of $\oint$ of biological systems.

We first consider a volatile environment where perturbations are added every time step ($O = 1$). Figure 5 shows that for this high level of noise, the network of *A. thaliana cell-cycle* is fragile, the networks of *death receptor signaling* and *tumor cell invasion and migration* are robust in a certain range of *X* and fragile in the rest of the range, and the networks of *CD4+ T cell differentiation and plasticity*, *mammalian cell-cycle*, *cardiac development*, and *metabolic interactions in the gut microbiome* are antifragile against perturbations. When comparing with Figure 3(a), we found that antifragility of the biological networks except for *A. thaliana cell-cycle* is similar to that of ordered or critical RBNs.

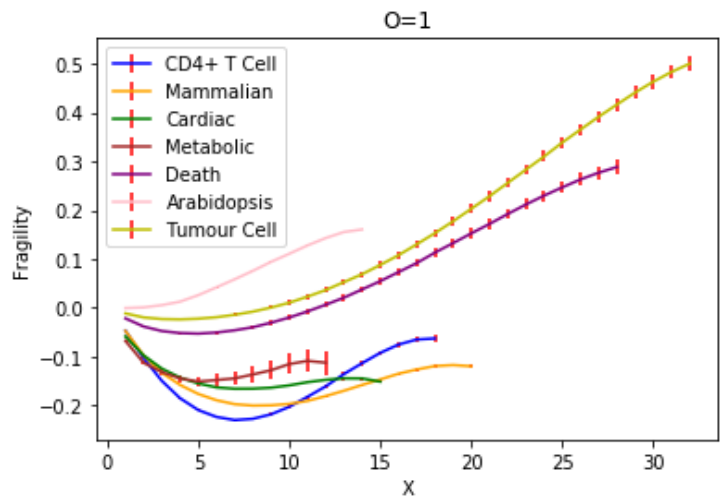

Figure 5: $\oint$ of biological Boolean networks. The error bars represent the standard error of measurements for 1000 different initial states run by 200 steps.

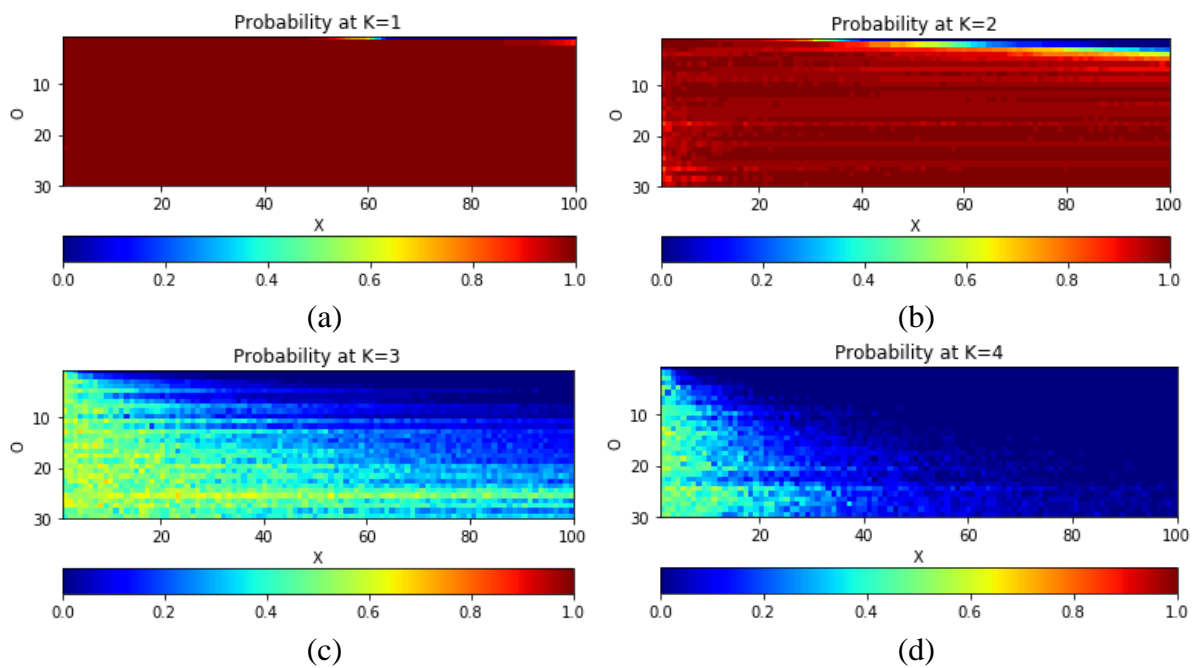

(a)

(b)

(c)

(d)

Figure 6: Probability of generating antifragile networks depending on *X* and *O* for ordered, critical, chaotic RBNs with $N = 100$, $T = 200$, $p = 0.5$. 50 different networks were used. 10 different initial states were randomly chosen for each network. (a) $K = 1$. (b) $K = 2$. (c) $K = 3$. (d) $K = 4$.

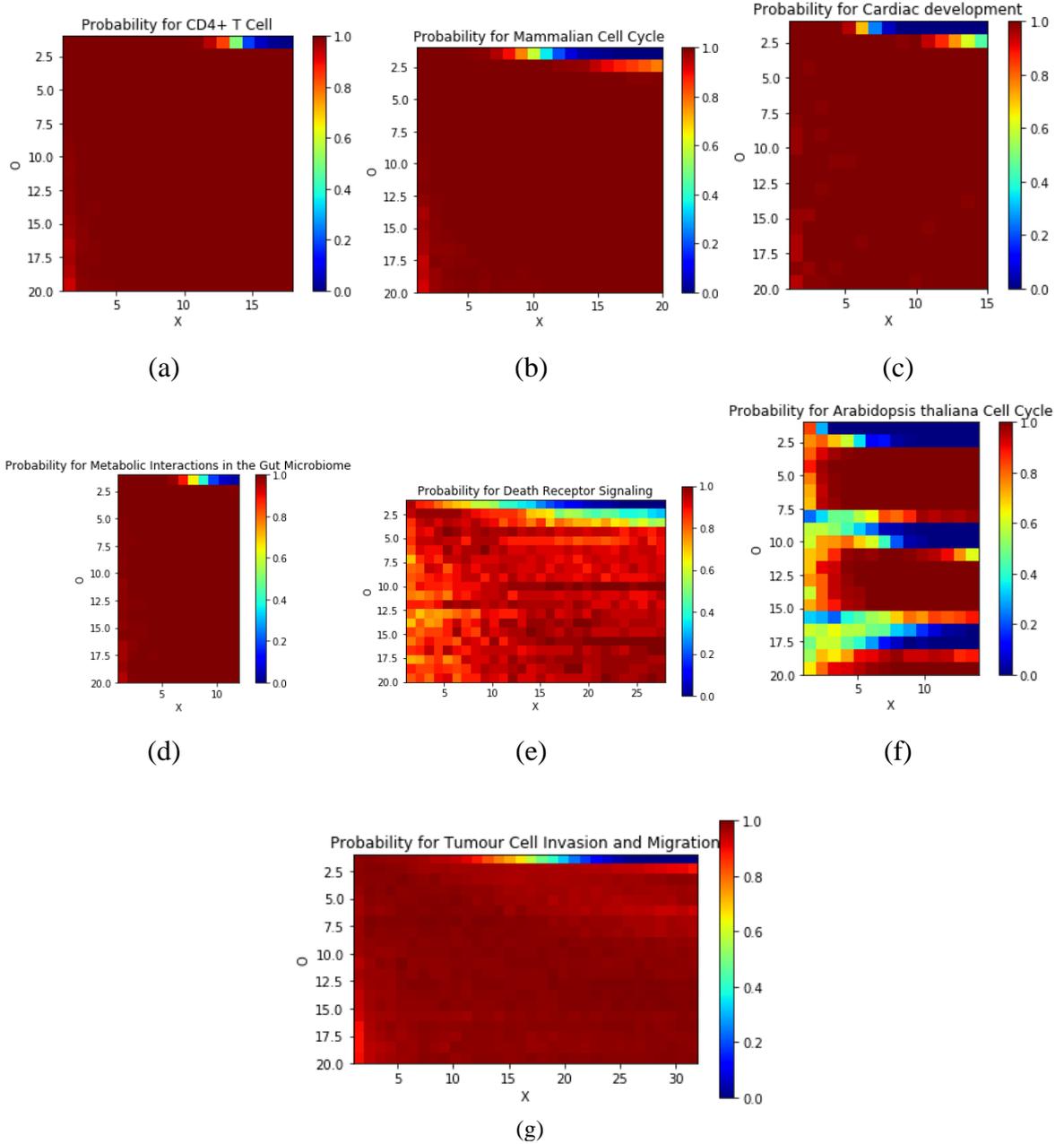

(a)

(b)

(c)

(d)

(e)

(f)

(g)

Figure 7: Probability of generating antifragile networks depending on *X* and *O* for different biological Boolean networks with $T = 200$. 5000 different initial states were used for each network. (a) *CD4+ T cell differentiation and plasticity*. (b) *Mammalian cell-cycle*. (c) *Cardiac development*. (d) *Metabolic interactions in the gut microbiome*. (e) *Death receptor signaling*. (f) *A. thaliana cell-cycle*. (g) *Tumor cell invasion and migration*.

To obtain more generalized dynamics, we investigated the probability of generating antifragile networks in a diverse range of *X* and *O*. Figure 6 is a heat map showing the probability for RBNs. As shown in the figure, the ordered and critical RBNs can produce antifragile networks. However, if too large perturbations are added in a volatile environment (*i.e.*, $O = 1$), both of them do not exhibit antifragile dynamics. In the case of the chaotic RBNs, they cannot produce antifragile networks in any range of *X* and *O*.

Figure 7 is a heat map for the seven BNs. They all show antifragile dynamics like the ordered or critical RBNs. Among the heat maps, the most interesting networks are *A. thaliana cell-cycle* and *CD4+ T cell differentiation and plasticity*. We found that *A. thaliana cell-cycle* repeatedly produces antifragile networks at regular intervals depending on the values of $O$. Based on many studies demonstrating living organisms are ordered or critical [47-50], we can infer that *A. thaliana* might have been evolved in environments where particular dimensions of perturbations are added more frequently than other biological systems. We also found that *CD4+ T cell differentiation and plasticity* is the most antifragile of the ones studied, probably because it has the most variable environment. It indicates that our antifragility measure successfully captures the property of the immune system mentioned as a representative example of antifragile systems.

## Conclusions

In this study, we proposed a new measure of (anti)fragility and applied it to RBNs. Considering an environment given to a system as a noise source, we observed how system properties can be varied depending on the degree of perturbations. We found that ordered and critical RBNs show antifragile dynamics, and especially ordered RBNs are most antifragile against the perturbations. Also, biological systems show antifragile dynamics.

In addition to the findings, we gained a meaningful insight to environments as external stressors. The high complexity with an optimal balance between regularity and change was acquired when a moderate perturbations were added very frequently. It means that "optimality" depends on the precise variability of the environment. How can systems be antifragile or robust for varying levels of noise? Which mechanisms can be used to adjust the internal variability depending on the external variability? These questions demand further studies, but possible answers are already being explored based on the results presented here.

Based on the findings and insight, by adjusting the size and frequency of perturbations, we can control system properties from fragile through robust to antifragile dynamics. It may help to understand dynamical behaviors of biological systems depending on environmental conditions and develop new treatment strategies for various diseases including cancer or AIDS, *e.g.* how can we decrease the antifragility of cancer cells or pathogens? This should reduce their adaptability and potentially improve treatments.

## Data Availability

Our simulator and data are available at https://github.com/Okarim1/RBN.git .

## Conflicts of Interest

The authors declare that there is no conflict of interest regarding the publication of this paper.

## Funding Statement

This research was partially supported by CONACYT and DGAPA, UNAM.

## Acknowledgments

We are grateful to Dario Alatorre, Ewan Colman, Luis Ángel Escobar, José Luis Mateos, Dante Pérez, and Fernanda Sánchez-Puig for useful comments and discussions.

I would like to acknowledge Dr. Gershenson for his guidance during the MSc course of Adaptive Computation. The final work of the course served as a foundation for this paper. [OKP]

# References

[1]   N. Taleb, *Antifragile : Things that gain from disorder*. Random House New York, 2012.

[2]   J. Derbyshire and G. Wright, "Preparing for the future: Development of an 'antifragile' methodology that complements scenario planning by omitting causation," *Technological Forecasting and Social Change*, vol. 82, pp. 215–225, 2014.

[3]   T. Aven, "The concept of antifragility and its implications for the practice of risk analysis," *Risk Analysis*, vol. 35, no. 3, pp. 476–83, 2015.

[4]   A. Naji, M. Ghodrat, H. Komaie-Moghaddam, and R. Podgornik, "Asymmetric coulomb fluids at randomly charged dielectric interfaces: Antifragility, overcharging and charge inversion," *The Journal of chemical physics*, vol. 141, no. 17, 174704, 2014.

[5]   M. Grube, L. Muggia, and C. Gostinčar, "Niches and adaptations of polyextremotolerant black fungi," *Polyextremophiles*, pp. 551–566, Springer, 2013.

[6]   A. Danchin, P. M. Binder, and S. Noria, "Antifragility and tinkering in biology (and in business) flexibility provides an efficient epigenetic way to manage risk," *Genes*, vol. 2, no. 4, pp. 998–1016, 2011.

[7]   J. S. Levin, S. P. Brodfuehrer, and W. M. Kroshl, "Detecting antifragile decisions and models lessons from a conceptual analysis model of service life extension of aging vehicles," *Proceedings of 2014 IEEE International Systems Conference*, pp. 285–292, 2014.

[8]   R. Isted, "The use of antifragility heuristics in transport planning," Proceedings of Australian Institute of Traffic Planning and Management (AITPM) National Conference, no. 3, 2014.

[9]   K. H. Jones, "Engineering antifragile systems: A change in design philosophy," *Procedia Computer Science*, vol. 32, pp. 870–875, 2014.

[10]  M. Lichtman, M. T. Vondal, T. C. Clancy, and J. H. Reed, "Antifragile communications," *IEEE Systems Journal*, vol. 12, no. 1, pp. 659–670, 2018.

[11]  E. Verhulst, "Applying systems and safety engineering principles for antifragility," *Procedia Computer Science*, vol. 32, pp. 842–849, 2014.

[12]  C. A. Ramirez and M. Itoh, "An initial approach towards the implementation of human error identification services for antifragile systems," *Proceedings of the SICE Annual Conference (SICE)*, pp. 2031–2036, 2014.

[13]  A. Abid, M. T. Khemakhem, S. Marzouk, et al., "Toward antifragile cloud computing infrastructures," *Procedia Computer Science*, vol. 32, pp. 850–855, 2014.

[14]  M. Monperrus, "Principles of antifragile software," Proceedings of Companion to the first International Conference on the Art, Science and Engineering of Programming, ACM, 32, 2017.

[15]  L. Guang, E. Nigussie, J. Plosila, and H. Tenhunen, "Positioning antifragility for clouds on public infrastructures," *Procedia Computer Science*, vol. 32, pp. 856–861, 2014.

[16] R. Serra, M. Villani, A. Barbieri, S. A. Kauffman, and A. Colacci, "On the dynamics of random Boolean networks subject to noise: attractors, ergodic sets and cell types," *Journal of Theoretical Biology*, vol. 265, no. 2, pp. 185–193, 2010.

[17] A. Bauer, T. L. Jackson, Y. Jiang, and T. Rohlf, "Receptor cross-talk in angiogenesis: mapping environmental cues to cell phenotype using a stochastic, Boolean signaling network model," *Journal of Theoretical Biology*, vol. 264, no. 3, pp. 838–846, 2010.

[18] M. K. Morris, J. Saez-Rodriguez, P. K. Sorger, and D. A. Lauffenburger, "Logic-based models for the analysis of cell signaling networks," *Biochemistry*, vol. 49, no. 15, pp. 3216–3224, 2010.

[19] R. Albert and H. G. Othmer, "The topology of the regulatory interactions predicts the expression pattern of the segment polarity genes in *Drosophila melanogaster*," *Journal of theoretical biology*, vol. 223, no. 1, pp. 1–18, 2003.

[20] F. Li, T. Long, Y. Lu, Q. Ouyang, and C. Tang, "The yeast cell-cycle network is robustly designed," *Proceedings of the National Academy of Sciences*, vol. 101, no. 14, pp. 4781–4786, 2004.

[21] M. Paczuski, K. E. Bassler, and A. Corral, "Self-organized networks of competing Boolean agents," *Physical Review Letters*, vol. 84, no. 14, 3185, 2000.

[22] M. E. Newman, "Spread of epidemic disease on networks," *Physical review E*, vol. 66, no. 1, 016128, 2002.

[23] P. Rämö, S. A. Kauffman, J. Kesseli, and O. Yli-Harja, "Measures for information propagation in Boolean networks," *Physica D: Nonlinear Phenomena*, vol. 227, no. 1, pp. 100–104, 2007.

[24] A. Roli, M. Manfroni, C. Pinciroli, and M. Birattari, "On the design of Boolean network robots," *Proceedings of European Conference on the Applications of Evolutionary Computation*, pp. 43–52, Springer, 2011.

[25] S. A. Kauffman, "Metabolic stability and epigenesis in randomly constructed genetic nets," *Journal of Theoretical Biology*, vol. 22, no. 3, pp. 437–467, 1969.

[26] S. A. Kauffman, The Origins of Order Self-Organization and Selection in Evolution. Oxford University Press, 1993.

[27] C. Gershenson, "Introduction to random Boolean networks," Workshop and Tutorial Proceedings of Ninth *International Conference on the Simulation and Synthesis of Living Systems (ALife IX)*, pp. 160–173, 2004.

[28] C. Gershenson and N. Fernández, "Complexity and information: Measuring emergence, self-organization, and homeostasis at multiple scales," *Complexity*, vol. 18, no. 2, pp. 29–44, 2012.

[29] C.C. Walker, W.R. Ashby, "On temporal characteristics of behavior in certain complex systems," *Kybernetik*, vol. 3, no. 2, pp.100–108, 1966

[30] B. Derrida and Y. Pomeau, "Random networks of automata: A simple annealed approximation," *Europhysics Letters*, vol. 1, no. 2, 45, 1986.

[31] C. Gershenson, "Guiding the self-organization of random Boolean networks, " *Theory in Biosciences*, vol. 131 no. 3 pp. 181–191, 2012.

[32] C. G. Langton, "Computation at the edge of chaos: phase transitions and emergent computation," Physica D: Nonlinear Phenomena, vol. 42, no. 1-3, pp. 12–37, 1990

[33] R. Lopez-Ruiz, H. L. Mancini, and X. Calbet, "A statistical measure of complexity," *Physics Letters A*, vol. 209, no. 5-6, pp. 321-326, 1995.

[34] G. Santamaría-Bonfil, C. Gershenson, and N. Fernández, "A package for measuring emergence, self-organization, and complexity based on Shannon entropy," *Frontiers in Robotics and AI*, vol. 4, 10, 2017.

[35] N. Fernández, C. Maldonado, and C. Gershenson, "Information measures of complexity, emergence, self-organization, homeostasis, and autopoiesis," In Prokopenko, M. (Ed.) *Guided self-organization: Inception*, pp. 19–51, Springer, 2014.

[36] C. Gershenson, "The sigma profile: A formal tool to study organization and its evolution at multiple scales," *Complexity*, vol. 16, no. 5, pp.37-44, 2011.

[37] M. E. Martinez-Sanchez, L. Mendoza, C. Villarreal, and E. R. Alvarez-Buylla, "A minimal regulatory network of extrinsic and intrinsic factors recovers observed patterns of CD4+ T cell differentiation and plasticity," *PLOS Computational Biology*, vol. 11, no. 6, 1004324, 2015.

[38] Ö. Sahin, H. Fröhlich, C. Löbke et al., "Modeling ERBB receptor-regulated G1/S transition to find novel targets for *de novo* trastuzumab resistance," *BMC Systems Biology*, vol. 3, no. 1, 1, 2009.

[39] F. Herrmann, A. Groß, D. Zhou, H. A. Kestler, and M. Kühl, "A Boolean model of the cardiac gene regulatory network determining first and second heart field identity," *PLOS ONE*, vol. 7, no. 10, 46798, 2012.

[40] S. N. Steinway, M. B. Biggs, T. P. Loughran, Jr, J. A. Papin, and R. Albert, "Inference of network dynamics and metabolic interactions in the gut microbiome," *PLOS Computational Biology*, vol. 11, no. 6, 1004338, 2015.

[41] L. Calzone, L. Tournier, S. Fourquet, et al., "Mathematical modelling of cell-fate decision in response to death receptor engagement," *PLOS Computational Biology*, vol. 6, no. 3, 1000702, 2010.

[42] E. Ortiz-Gutiérrez, K. García-Cruz, E. Azpeitia, et al., "A dynamic gene regulatory network model that recovers the cyclic behavior of *Arabidopsis thaliana* cell cycle," *PLOS Computational Biology*, vol. 11, no. 9, 1004486, 2015.

[43] D. P. A. Cohen, L. Martignetti, S. Robine, et al., "Mathematical modelling of molecular pathways enabling tumour cell invasion and migration," *PLOS Computational Biology*, vol. 11, no. 11, 1004571, 2015.

[44] C. Gershenson and D. Helbing, "When slower is faster," *Complexity*, vol. 21, no. 2, pp. 9–15, 2015.

[45] S. A. Kauffman, *Investigations*. Oxford University Press, 2000.

[46] J. Schuecker, S. Goedeke, and M. Helias, "Optimal sequence memory in driven random networks," *Physical Review X*, vol. 8, no. 4, 041029, 2018.

[47] I. Shmulevich, S. A. Kauffman, and M. Aldana, "Eukaryotic cells are dynamically ordered or critical but not chaotic," *Proceedings of the National Academy of Sciences*, vol. 102, no. 38, pp.13439–13444, 2005.

[48] E. Balleza, E. R. Alvarez-Buylla, A. Chaos, et al., "Critical dynamics in genetic regulatory networks: examples from four kingdoms," *PLoS One*, vol. 3, no. 6, 2456, 2008.

[49] M. Villani, L. La Rocca, S. A. Kauffman, and R. Serra, "Dynamical criticality in gene regulatory networks," *Complexity*, vol. 2018, 5980636, 2018.

[50] B. C. Daniels, H. Kim, D. Moore, et al., "Criticality distinguishes the ensemble of biological regulatory networks," *Physical Review Letters*, vol. 121, no. 13, 138102, 2018.